\UseRawInputEncoding
\documentclass[runningheads]{llncs}
\usepackage[T1]{fontenc}
\usepackage{graphicx}
\usepackage[%
  square,        
  comma,         
  numbers,       
  sort&compress  
]{natbib}

\usepackage{multicol}
\usepackage{multirow}
\usepackage{tikz}
\usepackage{ctable}
\usepackage{array}
\usepackage{makecell}

\usepackage{longtable}

\usepackage[inline,shortlabels]{enumitem}
\usepackage{csquotes}

\usepackage{hyperref}

\usepackage[utf8]{inputenc}

\begin{document}
\title{Architecturally Significant MLOps Guidelines for ML Model Integration and Deployment: a Gray Literature Review}
\titlerunning{Architecturally Significant MLOps Guidelines}
%
\author{Faezeh Amou Najafabadi\inst{1}\orcidID{0009-0006-6413-9118} \and
	Markus Haug\inst{2}\orcidID{0000-0001-9377-0677} \and
	Keerthiga Rajenthiram\inst{1}\orcidID{0009-0007-0885-5264} \and
	Justus Bogner\inst{1}\orcidID{0000-0001-5788-0991} \and
	Ilias Gerostathopoulos\inst{1}\orcidID{0000-0001-9333-7101}}
\authorrunning{F. Amou Najafabadi et al.}
%

%
\authorrunning{F. Amou Najafabadi et al.}
%
\institute{Vrije Universiteit Amsterdam, Amsterdam, The Netherlands \\
	\email{f.amou.najafabadi,k.rajenthiram,j.bogner,i.g.gerostathopoulos@vu.nl} \and
	Technical University of Munich, Munich, Germany \\
	\email{markus.haug@tum.de}}
\maketitle              

\begin{abstract}

\textit{Context}.
	Despite the growing adoption of Machine Learning Operations (MLOps), teams often approach MLOps projects in an ad hoc manner due to the lack of consolidated architectural guidance. The community would benefit from a reference that synthesizes knowledge to inform the architectural design of MLOps systems, especially regarding the integration and deployment of ML models.
	\noindent \textit{Objective}.
	In response, our goal is to provide a comprehensive overview of architecturally significant guidelines for the integration and deployment of ML models in MLOps systems.
	\noindent \textit{Method}.
	We conduct a gray literature review of 103 web sources to analyze state-of-practice knowledge on MLOps model integration and deployment. We then apply thematic analysis to synthesize these practices into recommended guidelines.
	\noindent \textit{Results}.
	We contribute a collection of 25 architecturally significant MLOps guidelines for model integration and deployment, organized into five categories, and describe their impact on the overall system architecture.
	\noindent \textit{Conclusion}.
	Our results serve as an overview of state-of-practice MLOps guidelines to support researchers and practitioners with the integration and deployment of ML models in their MLOps systems.

	\keywords{Machine Learning Operations \and MLOps \and MLOps Architecture \and MLOps Guideline \and ML Model Integration \and ML Model Deployment \and Gray Literature Review.}
\end{abstract}

\section{Introduction}

Engineering machine learning (ML) components is more than training and testing ML models; it also involves packaging and integrating them in ML-enabled systems, deploying them to production environments, monitoring them for data drift and performance degradation, and evolving them by online re-training without downtime. 
Realizing this, ML engineering teams are increasingly adopting best practices, methods, and tools that help them bridge the gap between development and operation of ML components --- a paradigm referred to as ML Operations (MLOps)~\cite{john_towards_2021,treveil_introducing_2021}.

Despite increasing understanding and broad adoption, operationalizing MLOps remains difficult for practitioners~\cite{amrit_analysis_2025,tamburri_challenges_2026}. 
First, being an emerging paradigm, there is no standard set of tools and related patterns. 
Although two predominant open-source stacks seem to emerge --- one around MLFlow, another around KubeFlow~\cite{berberi_machine_2025,micallef_systematic_2026} --- there are still several open architectural decisions even when choosing one of the two stacks as a starting point. 
Second, there is limited consolidation of MLOps architectural knowledge in terms of reusable design decisions, architectural tactics, or guidelines despite notable attempts in this direction~\cite{warnett_architectural_2022,warnett_bridging_2025}.
This makes it difficult to transfer the experience of an engineering team to new projects and to other teams. 
Third, MLOps has obtained a rather broad scope, covering both the initial phases of ML model conceptualization, requirements engineering, and model development, to the integration, deployment, monitoring, and operation of ML components~\cite{amou_najafabadi_architectural_2026}. 
This creates a large space of interconnected decisions that are difficult for any engineering team to take, even when the necessary architectural knowledge, experience, and tools are in place. 

To systematically support the application of MLOps practices, we need to understand how practitioners have been applying them (successfully) in their projects and derive the guiding principles from their collective experience. 
To promote reuse, we focus on \textit{architecturally-significant guidelines} (ASG), i.e., guidelines that have an impact on the MLOps system structure and key quality attributes such as scalability, maintainability, performance, security, or reliability. 
We define an MLOps system as a system composed of ML, non-ML, and infrastructure components that collect and process data; train and serve models; and monitor, manage, and retrain models~\cite{amou_najafabadi_architectural_2026}.
Our long-term goal is to incorporate ASGs into a holistic reference architecture for MLOps.

To keep our investigation focused, we only consider two key phases of MLOps: \textit{ML model integration} and \textit{deployment}. 
The former refers to the process of embedding ML models or ML artifacts alongside traditional software components~\cite{sens2024MLIntegration}. 
The latter refers to the process of integrating a model into a target environment so that it can be used to make predictions on real-world data~\cite{googleCloudArch}.
This paper addresses the research question:
\textit{What are common architecturally significant guidelines for integrating and deploying ML models in MLOps systems?}

We address this question by performing a systematic review of relevant gray literature, since it is a rich source of practices and guidelines for engineering MLOps.
We focus exclusively on gray literature since the topic of this review is a rapidly evolving, practice-driven domain in which peer-reviewed studies remain limited.
Our results contribute a taxonomy of five categories of architecturally significant MLOps guidelines comprising 25 guidelines.
The target audience of our study are (i) researchers in software architecture for MLOps, who can build on our results to derive and consolidate architectural guidelines for MLOps, and (ii) software and ML practitioners, who can use our findings to inform the design decisions for ML model integration and deployment in their MLOps systems.



The remainder of the paper is organized as follows. In Section~\ref{sec:related-work}, we discuss the related work and position our contribution. Section~\ref{sec:methodology} describes our research method, while we present the results of the study in Section~\ref{sec:results}. The results with their implications are discussed in Section ~\ref{sec:discussion}, while the threats to validity in Section~\ref{sec:threats}. Finally, Section~\ref{sec:conclusion} concludes this paper and discusses future work.

\section{Related work} \label{sec:related-work}


With the increasing importance and adoption of ML-enabled systems in practice, researchers are investigating challenges and practices for the successful engineering of these systems.
One of the first studies to investigate software engineering practices for ML-enabled systems was provided by~\citet{Washizaki2019}.
They conducted a multivocal literature review to identify design patterns for ML-enabled systems.
However, their study was conducted in the early days of ML-enabled systems, before MLOps became widely known.
As such, the best practices may have changed since their study was published, and a new investigation is worthwhile.

Based on the work of \citet{Washizaki2019}, \citet{warnett_architectural_2022} combined a GLR with grounded theory to synthesize architectural design decisions for machine learning deployment.
Their scope exclusively addressed design decisions for the deployment of ML components, while our study focuses on MLOps and additionally covers practices for ML model and component integration.
\citet{warnett_architectural_2022} also identified the adoption of MLOps as an important decision in their study, but did not fully cover how MLOps shaped related practices.

\citet{Heiland2023} presented a multivocal literature review identifying design patterns for AI-based systems substantially extending the scope of \citet{Washizaki2019}.
In contrast to our study, they investigated more general patterns for AI-based systems without a clear MLOps focus.
They also developed a web-based platform to help practitioners explore their pattern catalog.

\citet{Nahar2023} contributed a meta summary of challenges in developing ML-enabled systems.
They analyzed 50 prior studies, interviewing or surveying over 4758 participants, to identify a comprehensive set of challenges in building software products with ML components.
In contrast to our study, they did not incorporate gray literature in their study. Additionally, unlike our study, they report on the challenges without providing guidelines to address them.

\citet{Alves2023} conducted a multivocal literature review of practices for managing ML products, \cite{eken2025multivocal} performed a multivocal literature review of MLOps practices and challenges, and \citet{zarour2025mlops} conducted a systematic literature review of MLOps best practices, challenges, and maturity levels.
In contrast to our study, the resulting practices of these studies are described as more high-level guidelines.
These studies also focus broadly on the whole ML lifecycle, whereas our study investigates the integration and deployment phases in detail and from an architectural perspective.

\citet{kumara2025mlops} contributes a reference architecture with a focus on requirements for an MLOps environment by reviewing gray literature. Similar to \cite{Alves2023} and \cite{zarour2025mlops}, the authors take a holistic view of MLOps, and as opposed to our study, they prioritize reporting on MLOps requirements and related tools without providing guidelines for any stage of MLOps.

Overall, while existing work has shown practitioner-oriented literature to be a valuable resource for research into best practices in engineering MLOps and ML-enabled systems, to our knowledge, no prior research has investigated best practices to provide guidelines for the integration between ML components and conventional parts of a software system in detail.
Our study aims to fill this gap.

\section{Methodology} \label{sec:methodology}
In this research, we follow the guidelines for conducting gray literature reviews (GLR) specified by Garousi et al. in \cite{garousi2019_guidelinesMLR}. 
In the following, we describe the process and our approach in conducting this GLR.
For transparency and reproducibility, we provide all our study materials in a replication package~\cite{replication_package}.

\subsection{Goal and research question} \label{subsec_Goal-RQ}

The main aim of this GLR is to synthesize architecturally significant guidelines for ML model integration and deployment in MLOps systems. 
The results of the study are aimed at researchers and practitioners who intend to build upon community experience when integrating and deploying ML models.
To approach this goal, we address the following research question:
%
%
\textbf{What are common architecturally significant guidelines for integrating and deploying ML models in MLOps systems?}
This research question aims to capture how practitioners integrate and deploy ML models, what they do in this regard, and which practices they advise to follow or avoid.


\subsection{Research process}

Here, we describe the different steps of our research process. Essentially, we first obtain an initial set of sources via an automated search via the Google search engine and save the retrieved sources for further steps. After finalizing the source list, we extract data from the dataset of retrieved sources. Finally, the extracted data, which is in open-format text, is used in an iterative synthesis process.
\subsubsection{Step 1: Automated initial search} 

Considering the goal and the research question, we formulate the following query and use it in the Google search engine:

\small
\noindent\textit{MLOps AND Architect AND (``Model Integration'' OR Deployment OR CI/CD) AND (Practice OR Guideline OR Tactic OR Strategy OR Recommendation OR Pattern OR ``Best Practice'' OR Antipattern OR Smell OR ``Technical Debt'' OR Pitfall)}
\normalsize

Essentially, the query consists of four parts combined with \textit{AND}: MLOPs, architecture (we used the stem ``architect''), MLOps phase (three keywords), and practice/guideline (11 keywords). For the first part, although we initially considered using the full term ``Machine Learning Operations'', we decided against this since in our selection pilot we found that almost all the sources use the abbreviation ``MLOps''. For the last part, we use several synonyms to capture most if not all of the different terms used by practitioners to refer to good/best/bad practices and recommendations. In our search strategy, we used quotation marks only for multi-word constructs, and left single-word keywords unquoted to avoid overly restrictive exact-match behavior in Google and to balance specificity with search coverage.


To search on Google, we split the above query into several simple queries since the search engine does not handle complex strings well. 
Specifically, we construct all the simple queries derived by combining an ``MLOps phase'' keyword with a ``practice'' keyword. For example, one of the simple queries was \textit{MLOps Architect “Model Integration” Practice.}; another was \textit{MLOps Architect “Model Integration” Guideline.}, etc.
We also use \textit{incognito mode} to avoid search bias produced by personalized search results.

In the end, this resulted in 33 simpler queries that are run on Google, and the outcomes of the first two pages of each simpler query are considered. 
This stopping criterion of two pages per query is based on the search engine page rank algorithm~\cite{longville2006_GooglesPageRank} according to the GLR guidelines in~\cite{garousi2019_guidelinesMLR}. Additionally, our search pilot revealed substantial overlap among results appearing after the second page.

After collecting the results from all queries and removing duplicates, we save 331 (web) sources to our dataset for further steps.

\subsubsection{Step 2: Source selection} To select the sources for data extraction, we apply the inclusion (I) and exclusion (E) criteria as follows:

\begin{itemize}
    \item[\textbf{I1}] The source describes at least one practice (good or bad) in an MLOps system.
    \item[\textbf{I2}] The described practice(s) are architecturally significant, i.e. they
    have an impact on the system structure and key quality attributes such as scalability, maintainability, performance, and reliability.
    We define such practices by analogy of \textit{architecturally significant requirements} by Bass et al.~\cite{bass2012_software-architecture}.
    \item[\textbf{I3}] The described practice(s) are related to ML model integration or model deployment in the MLOps system.
    \item[\textbf{E1}] The source is in a language other than English.
    \item[\textbf{E2}] The source is an advertisement or introduction to another source, e.g., a full course or a commercial tool.
    \item[\textbf{E3}] The source is in a multimedia format, e.g., videos.
    \item[\textbf{E4}] The source is a scientific paper or publication rather than being practitioner-oriented.	
    \item[\textbf{E5}] The source is not freely available or accessible to everyone.
\end{itemize}

We start the selection process with a selection pilot on 5\% of the retrieved sources (i.e., 17 sources). In this step, all the authors individually apply the I/E criteria. The authors then discuss the selection process in a consensus meeting to clarify any doubts and clarify the definitions. 
Following the pilot and according to the I/E criteria, each source is examined by two researchers to assess whether it is included in the study. The selection decisions are then discussed in consensus meetings among all authors. Eventually, 103 sources are selected for data extraction.

\subsubsection{Step 3: Data Extraction} 

In this step, we extract the relevant text chunks referring to MLOps practices from each of the selected sources. 
Each source is extracted by two authors, and the final results are discussed in consensus meetings among all authors. 
To align our understanding on the abstraction level and scope of our extractions, we performed an extraction pilot on 24 sources (i.e., almost 25\% of the sources). The identification and discussion of disagreements and ambiguities in the extraction process during this pilot led to refined extraction guidelines that were subsequently applied by all authors to the remaining sources.



\subsubsection{Step 4: Data analysis and synthesis}
To synthesize the extracted practices, we perform thematic analysis on unstructured data by following a three-step approach. 
First, each extracted practice is assigned to three authors who independently synthesize it into an MLOps guideline. 
We define an MLOps guideline as a practical piece of advice on how practitioners should design, develop, and maintain MLOps systems. 
Due to the unstructured nature of the data, extensive meetings are held to agree upon the formulation of each guideline. 
In the second step, one author further refines and organizes these guidelines into a coherent list. The resulting list of guidelines is then distributed among the other authors for a final validation review.
Lastly, we perform open card sorting~\cite{Zimmermann2016_DS4SE} on the final list of synthesized guidelines to organize them into emerging categories.





As part of the methodology and in line with the qualitative nature of the analysis and the inductive categorization process, we adopted an iterative refinement strategy throughout the study.
In the initial stages, we took a more inclusive approach with regard to guidelines and categories. However, in a final synthesis iteration, we took a more focused approach and removed:

\begin{itemize}
    \item Categories that were not substantially enough related to our envisioned scope of deployment and integration. Two categories were removed, \textit{Data Management} and \textit{Process and Stakeholders}, with a total of 16 guidelines.
    These guidelines only indirectly affect the decisions made in the integration and deployment phases, so we consider them less central to our study.
    \item Guidelines in the remaining categories that did not appear frequently enough in our sources. We made this decision based on the \enquote{Rule of Three} from design pattern mining~\cite{rule-of-three}, leading to the removal of 12 guidelines.
\end{itemize}


\section{Results} \label{sec:results}

Based on 103 gray literature sources, our thematic analysis and subsequent refinements resulted in the 25 unique guidelines shown in Table~\ref{tab:guidelines_selected}. We have organized them into five categories that indicate different concerns related to model integration and deployment in MLOps from an architectural perspective. 
%
As seen in the table,
72\% of the guidelines (18/25) are mentioned more than four times, with some of them mentioned several times (e.g., A1, E1). This is an indication of the popularity of many of the practices among practitioners.
%
Due to space limitations, in the rest of the section, we describe the top two most frequently reported guidelines of each category. For the description of the rest of the guidelines, refer to the replication package~\cite{replication_package}.

\begin{table}[!t]
    \centering
    \scriptsize
    \caption{The list of architecturally significant MLOps guidelines per category with corresponding counts of supporting gray literature sources.} \label{tab:guidelines_selected}
    \begin{tabular}{|m{0.046\textwidth}|m{0.038\textwidth}|m{0.835\textwidth}|m{0.035\textwidth}|}
    \hline
        ~ & \textbf{ID} & \textbf{Guideline} & \textbf{\#} \\ \hline
         \parbox{25mm}{\multirow{6}{*} {\rotatebox[origin=r]{90}{ \makecell{CI/CD and \\ automation }}}}  
         & 
         A1 & Establish mature CI/CD pipelines for ML components. & 53 \\ \cline{2-4}
        ~ & 
        A2 & Make extensive use of automation for recurring tasks in the MLOps process. & 21 \\ \cline{2-4}
        ~ & 
        A3 & Design for continuous experimentation and small, frequent updates. & 11 \\ \cline{2-4}
        ~ & 
        A4 & Treat model training as an automated build job in the CI/CD pipeline. & 11 \\ \cline{2-4}
        ~ & 
        A5 & Set up automatic rollback mechanisms if a new ML component performs worse than the old one. & 9 \\ \cline{2-4}
        ~ & 
        A6 & Use human-in-the-loop gated approval for deploying models to production. & 6 \\ \hline
        
        \parbox{25mm}{\multirow{8}{*} {\rotatebox[origin=r]{90}{ \makecell{Deployment strategies \\ and environments }}}}   
        & 
        B1 &  Carefully select your production deployment strategy according to your context. & 18 \\ \cline{2-4}
        ~ & 
        B2 & Choose an appropriate type of deployment topology depending on the organizational context. & 12 \\ \cline{2-4}
        ~ & 
        B3 & Deploy models to a staging environment before deploying to production. & 7 \\ \cline{2-4}
        ~ & 
       B4 & Deploy the entire ML training pipeline as part of the CI/CD process rather than deploying only the trained models. & 4 \\ \cline{2-4}
        ~ & 
       B5 & Use a UAT (User Acceptance Testing) environment for code transition from dev-test to production environment. & 3 \\ \cline{2-4}
        ~ & 
       B6 & Use separate accounts per environment (development, staging, production). & 3  \vspace{4pt} \\ \hline
        
        \parbox{25mm}{\multirow{-1.5}{*} {\rotatebox[origin=r]{90}{ \makecell{Design and \\ integration strateg. }}}}
         & 
         \vspace{4pt} C1 & Adopt an architectural design principle or architectural style that suits the system’s requirements. & 13 \\ \cline{2-4}
        ~ & 
        \vspace{4pt} C2 & Separate the training pipeline from the deployment pipeline (e.g., via a model registry) & 5 \\ \cline{2-4}
        ~ & 
       \vspace{4pt} C3 & Use standardized model export formats for increased portability (ONNX, PMML, etc.). & 5 \\ \cline{2-4}
        ~ & 
        \vspace{6pt} C4& Use a model registry for improved reproducibility and governance. & 4  \vspace{10pt}
        %

       \\ 
       
       \hline
        \parbox{25mm}{\multirow{6}{*} {\rotatebox[origin=r]{90}{\makecell{ Model serving \\ and inference }}}}
         & 
        D1 & Design or choose an appropriate component to embed your ML model. & 24 \\ \cline{2-4}
        ~ & 
        D2& Choose an appropriate inference mode for your use case. & 8 \\ \cline{2-4}
        ~ &  
        D3& Consider multi-model endpoints to serve multiple ML models or components. & 4 \\ \cline{2-4}
        ~ & 
       D4 & Optimize costs for ML model serving infrastructure while maintaining acceptable quality and reliability. & 3 \\ \cline{2-4}
        ~ & 
       D5 & Reduce model size (compression, quantization) for edge deployment. & 3 \\ \hline

        \parbox{25mm}{\multirow{2}{*} {\rotatebox[origin=r]{90}{ \makecell{ML component \\ management }}}}
         &  
         E1 & Use containerization for ML component deployment and management. & 27 \vspace{2pt} \\ \cline{2-4}
        ~ &  
        E2& Use a container orchestration tool to manage the different MLOps system components. & 14 \\ \cline{2-4}
        ~ & 
        E3& Provision sufficient compute resources for ML components. & 14 \vspace{2pt} \\ \cline{2-4}
        ~ & 
       E4 & Distribute the workload across multiple interconnected compute nodes. & 3 \vspace{2pt} \\ \hline
    \end{tabular}
\end{table}

\subsubsection{CI/CD and Automation.} 
This category gathers guidelines related to continuous engineering and automation. 

\textbf{A1 - Establish mature CI/CD pipelines for ML components.} CI/CD pipelines streamline the development and deployment process by automating the build, test, and deployment phases of ML components. Implementing CI/CD pipelines enhances consistency and efficiency across MLOps systems, accelerates delivery cycles, and allows teams to bring innovations to market more quickly by responding promptly to evolving requirements while ensuring greater confidence in the reliability of their ML solutions. The automation via CI/CD pipelines reduces the risk of human error and enhances the overall reliability of MLOps systems (S9, S19). To keep things safe, it is suggested to integrate automated ML model quality assurance steps into the CI/CD pipeline. For instance, S217 mentions that ``CI/CD pipelines should automatically execute test suites to make sure the newest version is stable and ready for deployment''. While fully automated CI/CD is recommended, it is not always necessary (S172). In such cases, it is suggested to consider using a more manual, script-driven ML deployment workflow if the models are expected to rarely change in production, which is equivalent to Microsoft's MLOps maturity level 0~\cite{microsoft-maturity-model}. On the contrary, the adoption of fully mature CI/CD pipelines is equivalent to Microsoft's MLOps maturity level 4 (S1). 
To take a moderate approach and when teams do not intend to experiment with many different ML models, they are advised to use automatic ML pipelines, without adopting a fully mature CI/CD pipeline (S172). 
In this approach, teams are advised to deploy new models based primarily on new data rather than new code.
Moreover, it is often observed that many teams leave the subject of deployment until very late in their MLOps project, resulting in many surprises and challenges at the time of deployment. For example, discovering that the data used for training is not available in production. To avoid such challenges and obtain a well-structured MLOps system, teams are advised to start new projects with a full CI/CD pipeline (S316, S327).

\textbf{A2 - Make extensive use of automation for recurring tasks in the MLOps process.} 
This guideline is complementary to establishing mature CI/CD pipelines. However, it is more general in scope since it pertains not only to automating the steps related to the release of new ML models/components, but also automating the steps related to the evolution of deployed models via e.g., automated monitoring and re-training.
On the spectrum of fully manual to fully automated processes in MLOps systems, manual systems are advised for small MLOps projects when the team requires an easy approach for a start, with no typical specialized MLOps or architectural knowledge. However, such manual systems can easily break and are not suitable for large-scale requirements. In addition, the missing automation in such systems makes it very hard to train and deploy models that can adapt to changes fast enough (S172). To develop systems that adapt to frequent changes and reduce human effort in large-scale settings, organizations are advised to utilize consistent and fully automated pipelines for model development, testing, deployment, release, monitoring, and re-training. Designing and building a system to perform the tasks automatically reduces the likelihood of manual errors, ensures reproducibility across different stages of the model lifecycle (S15, S596), and is integral to maintaining agility, allowing rapid iteration and updates in response to evolving requirements (S19).

\subsubsection{Deployment strategies and environments.} 
This category contains guidelines related to when, where, and how to deploy different artifacts (data, code, models, containers). 

\textbf{B1 - Carefully select your production deployment strategy according to your context.} 
By deployment strategy, we refer to the way new ML components are deployed to production environments and made available to end-users. 
Deployment strategies include \textit{shadow deployment} (S217), \textit{canary deployment} (S284), \textit{A/B testing} (S246, S284), and \textit{blue/green deployment} -- optionally with gradual traffic shifting (S210). When selecting one of these strategies, important dimensions include downtime during deployment, condition-based deployment (e.g., automatic checks that determine the progression of the rollback), rollback time in case of failure, and additional deployment costs (e.g., associated with blue/green deployment). S246 mentions e.g., that ``by carefully weighing each of these dimensions, organizations can select the optimal deployment strategy for their needs, ensuring a smooth and efficient deployment process with minimal risk and downtime''. 
A typical trajectory is to progress from shadow deployments (lower effort) to canary deployments and then A/B testing -- and use blue/green deployment only if the infrastructure cost is manageable by the organization (since one needs to maintain two separate production environments in this strategy). 

\textbf{B2 - Choose an appropriate type of deployment topology depending on the organizational context.} An important decision is \textit{where} the ML components will be deployed and run. Among the possible deployment environments, we distinguish between: (1) cloud-based deployment using managed cloud services or dedicated servers, (2) on-premises deployment, (3) hybrid approach, (4) deploying at the edge. 
The choice depends on the specific organizational needs and goals, considering factors such as data privacy requirements, existing infrastructure, and budget constraints (S198).
\textit{Cloud services} provide the infrastructure and resources needed for hosting models, ensuring reliable performance, scalability, easy maintenance, and are suited for models requiring significant computational power and flexibility but with potential latency issues (S210). 
S211 provides clear guidance: ``In case of no strict data privacy requirements and regulations, cloud-based solutions are a good choice due to the unlimited infrastructure resources for model training and serving. 
%
An \textit{on-premises} solution would be acceptable for very strict security requirements or if the infrastructure is already available within the company.
The \textit{hybrid} solution is an option for companies that already have part of the systems built but want to extend them with additional services — e.g., to buy a pretrained model and integrate with the locally stored data or incorporate into an existing business process model''.
Finally, regarding edge deployments, S210 mentions that ``\textit{Edge} deployment involves running models directly on devices like smartphones or IoT devices, which reduces latency and improves data privacy but can be limited by computational resources. 
It is ideal for real-time processing applications, while cloud deployment is better suited for models requiring significant computational power and flexibility''.

\subsubsection{Design and integration strategies} 
This category groups guidelines related to the design of the different pipelines and their interactions (e.g., training, deployment) and the related architectural styles and principles.

\textbf{C1 - Adopt an architectural design principle or architectural style that suits the system’s requirements.} The best suited architectural style should be chosen considering the organization's goals and capabilities. Traditionally, AI systems were developed as monolithic applications where all components—from data processing to model inference—are contained within a single codebase. This approach can simplify initial development and can be used for simple systems. However, it struggles with scalability, maintainability, and agility in production environments (S390). To address these issues in modern MLOps systems, modular and serverless architectures are often advised. Modularity supports easier troubleshooting and development (S187). Its adoption addresses ``Glue Code'' or ``Pipeline Jungles'' which are the result of unplanned messy development and prevent effective and efficient model integration in the production system by a modular design with clear interfaces and separation of concerns (S194). As put in S205, ``Using a loosely coupled architecture also affects the extent to which a team can test and deploy their applications on demand, without requiring orchestration with other services''. In modular architectures, the advice is to make your interfaces specific and not to define and build large, messy, generic interfaces that interact with unnecessary components (S310). Serverless architectures are also mostly advised in case automatic scaling based on real time demand is needed (S210). Serverless deployments allow you to focus on writing source code that actually delivers business value, leaving the platform to handle complicated hardware and scaling questions (S217). For time-sensitive ML applications, the Lambda architecture is recommended, which combines batch and real-time processing to handle large-scale data ingestion, processing, and analytics. (S229)

\textbf{C2 - Separate the training pipeline from the deployment pipeline (e.g., via a model registry).}
The model training code and the model deployment code should be managed separately, and an update to one should not necessarily require an update to the other (S284). This provides an opportunity to update the model configuration and infrastructure without affecting the training workflow (S448). 
S5 lists several key reasons for such separation:
\begin{enumerate}
    \item ``Complexity and resource requirements: Model training often demands significant computational resources, including specialized hardware like GPUs. Integrating such resource-intensive tasks into the deployment phase is impractical and could impede the efficiency of the code integration process.
    \item Separation of concerns: By decoupling the model training phase from the deployment, there is greater flexibility in the development workflow. Training can be conducted independently with various parameters and evaluation methods. Simultaneously, deployment pipeline can proceed unencumbered, enabling direct deployment of the model using pre-existing trained models without necessitating retraining during each integration and deployment cycle.
    \item Iterative nature of ML development: Model training in ML is inherently iterative, often involving experimentation with different parameters and methods. combining this iterative training with deployment would significantly slow down the process, hindering the rapid iteration and integration that CI aims to achieve.''
\end{enumerate}

\subsubsection{Model serving and inference} 
The guidelines within this category represent best practices regarding the methods and approaches for serving trained and validated ML models in production and the actions and processes for inference.

\textbf{D1 - Design or choose an appropriate component to embed your ML model.} 
The ML model may be embedded in an appropriate component to expose its predictive functionality in several ways: 
\begin{enumerate}
    \item \textit{Embed the model directly into the component/service that is using it}, which is advisable when the model is consumed by a single business service and the inference pipeline is not so complex or costly. This simple approach can work well when models do not need to be retrained very often. The downside is that your application must be released each time the model is changed (S217, S316). 
    \item \textit{Deploy a dedicated service that hosts the model}, which should be used when the model is reused by business services and its transformation logic is not so complex or costly. This approach improves flexibility, maintainability, and ease of deployment (S251, S316).
    \item \textit{Break the inference pipeline into several services}, which is advisable when the inference pipeline is complex and/or with expensive computations re-usable from different models. 
    This approach enables the update or scaling of individual pipeline steps (e.g., pre-processing), fostering faster innovation cycles and more resilient deployments (S390).
    \item \textit{Use models published as data}, where, as put in S217, ``the model is retrieved as a data artifact, usually directly from a model registry. This allows the model to be updated on the fly without stopping or redeploying your service. It also enables advanced strategies such as Blue/Green and canary deployments. In this case, updates to the model can be scheduled periodically, or triggered by an API call which causes the API to retrieve a new model artifact''. 
\end{enumerate}
In the above approaches, emphasis should be put on wrapping the model into a RESTful API during packaging (S203, S206).

\textbf{D2 - Choose an appropriate inference mode for your use case.} In general, the ML component can provide inference in two different modes: batch serving or online serving (also called real-time serving). Batch serving is advised for high-throughput use cases where accurate predictions are required, e.g., sales and inventory analysis, log analysis and anomaly detection, and credit scoring and loan approval (S198, S265, S406). In the case of LLM inference, continuous batching across sequences is also recommended instead of static batching. Continuous batching allows the engine to dynamically and asynchronously batch tokens across sequences and users, without requiring synchronized request timing (S91). On the other hand, online serving is advised for low-latency use cases (S265), particularly for applications where quick and immediate responses are required based on incoming data, e.g., fraud-detection, chatbots, recommendation systems, and IoT applications (S198).

\subsubsection{ML component management} 
The guidelines in this category aim to help practitioners create ML components (ML models along with their dependencies) and manage ML models and ML components reliably and efficiently.

\textbf{E1 - Use containerization for ML component deployment and management.} 
In model serving, containerization helps to simplify the deployment process by packaging models and their dependencies into a container that can be easily deployed across different production environments (S26, S38). This approach ensures consistency, portability, and reproducibility between different environments and also reduces the risk of dependency conflicts (S4, S5). In case of using containerization, the deployment workflow pushes the container image to a container registry. S5 notes that ``This could be DockerHub, Amazon's ECR, Google's GCR, or another similar service. The choice of registry often depends on the deployment platform and organizational preferences''. Docker is the de facto tool most commonly mentioned in containerization across sources (S13).
Serialization (e.g., via the pickle library) also offers a lightweight way to package model artifacts (S84), though it provides fewer compatibility guarantees than containerization.

\textbf{E2 - Use a container orchestration tool to manage the different MLOps system components.} 
To guarantee that models can manage fluctuating demands, orchestration tools like Kubernetes are suggested to coordinate scalable deployments (S187). Orchestration tools are specifically essential for larger-scale systems to manage multiple containers across different servers, automatically handling load balancing, scaling, and failover (S84, S199). In particular, S231 explains that ``Orchestration automates and optimizes container management, leading to streamlined operations, scalability, and reliability in modern software development'', while S203 notes that ``Making the entire system portable through Kubernetes will also ensure that we can easily move from an on-premise solution to a Cloud or hybrid solution if the need arises''.

\section{Discussion} \label{sec:discussion}
In this section, we compile noteworthy takeaways and study reflections.
The scope of our study was integration and deployment within MLOps systems.
When looking at the thematic distribution of the extracted fragments and synthesized guidelines, we see that most MLOps practitioner sources focus more on the deployment than the integration part.
For example, categories closer to deployment, like \textit{CI/CD and automation}, \textit{Deployment strategies and environments}, and \textit{ML component management} had a joint average of 13.5 sources per guideline (median of 11).
Conversely, categories more closely related to integration, like \textit{Design and integration strategies} and \textit{Model serving and inference}, had a joint average of only 7.66 sources per guideline (median of 5). 
Additionally, the number of guidelines in the deployment-related categories was considerably larger than in the integration-related ones (16 vs. 9).
This points to a difference in maturity regarding reusable architectural guidance in these two important phases, with the integration phase much less documented.
Considering that previous research has identified a strong impact of integration-related decisions like ML model export formats~\cite{KumarParida2025} as well as substantial integration heterogeneity in ML open-source projects~\cite{sens2024MLIntegration}, this lack of practitioner guidance is concerning.
However, this is also an opportunity for the AI engineering research community to address this gap and to provide reusable guidance in this space.

One thing we observed during the source analysis was that many practitioners recommended using specific tools instead of more generalizable practices.
For example, sources suggested concrete CI/CD tools like Jenkins or GitHub Actions, Kubernetes as a container orchestration tool, or specific holistic ML cloud platforms like Amazon SageMaker or Azure Machine Learning.
Often, these recommendations were not sufficiently justified in terms of expected benefits or a more generalizable reason behind the use of the tool.
As a general rule, we only used such tool-focused extractions for guideline synthesis if the concrete tool name could be omitted without losing generalizable usefulness.

Additionally, many sources described their recommended practices very briefly and abstractly, sometimes with as little as one sentence or bullet point, which may hinder their application.
For guidelines that are based on many sources, we were still able to synthesize a more thorough description from the combined related extractions.
In some cases, however, this led to descriptions that are not as actionable as those of other guidelines and thus remain largely descriptive.

Another takeaway from this study is that MLOps practitioner sources use related terminology very inconsistently, with sometimes very different scopes and meanings.
A prime example of such an overloaded term was \enquote{deploying}, which was used to mean anything from making models or components available in a computing environment for their usage (the meaning we use in this paper) to integrating a model into a component or even serving a model for inference in a specific way.
This made it challenging to collocate extractions and to disambiguate similar-sounding practices.
We hope that our synthesis and categories can bring a level of joint understanding to this space.

Moreover, our search string included both terms for good practices and bad practices.
Initially, we planned to report separately on those, but it turned out that MLOps practitioners mostly write about good practices, i.e., what to do, and not so much about bad practices, i.e., what not to do.
We therefore decided to phrase all our synthesized guidelines as good practices by inverting a potential bad practice.
For example, an extraction of the form \enquote{There's a common anti-pattern I see in software projects [...]. Leaving the deployment of your application until there's something meaningful to deploy.} (S327) became part of the \enquote{Start a new ML project by creating the CI/CD pipeline} guideline.\footnote{Note: this guideline was in the end merged into A1 as a sub-guideline.}

Lastly, our study focused mainly on the curation of architecturally significant MLOps guidelines, their description, and categorization.
However, applying several guidelines may be subject to challenges and trade-offs depending on the system context.
Although most of the reviewed practitioner sources did not mention such factors, understanding them better is important for the effectiveness of our guideline collection.
Practitioners using the guidelines need to be enabled to understand suitable system contexts as well as to navigate potential trade-offs.
We will address this in a future extension and evaluation study, which will assess the understandability, relevance, significance, and usefulness of the synthesized guidelines and validate the proposed taxonomy through focus groups with MLOps stakeholders. 

\section{Threats to validity} \label{sec:threats}
This GLR is subject to several threats to validity related to search, selection, and the nature of web-based sources. A key threat concerns search and source selection. We relied primarily on Google; while this may limit coverage by excluding sources indexed elsewhere, we selected Google due to its broad coverage and frequent use in GLRs. We also limited Google screening to the first two result pages, which may introduce ranking bias; to reduce personalization effects, searches were conducted in incognito mode.
Another threat arises from the dynamic nature of web sources, as webpages may change or become unavailable over time. We mitigated this by systematically extracting and recording relevant data at the time of access using a structured extraction form, although this limitation cannot be fully removed. A further threat is researcher subjectivity in source selection and data extraction, especially when judging whether a practice is architecturally significant and relevant to ML model integration or deployment. To address this, we defined explicit inclusion and exclusion criteria, had two (and in some cases three) researchers independently assess each source, and resolved disagreements through consensus meetings. In this setting, one researcher was involved in all assessments to ensure consistency and reduce the risk of fragmented overview. We also used a structured coding process to reduce interpretation bias.
Finally, coverage and source quality remain potential threats, since gray literature is heterogeneous and uneven in credibility, and can even contain replicated content generated by AI. To mitigate these, we excluded advertisements, inaccessible sources, and non-text formats, and we also relied on the extraction and assessment of more than one reviewer. Overall, although these threats cannot be fully eliminated, our systematic process improves the reliability and transparency of the findings.

\section{Conclusion} \label{sec:conclusion}
In this paper, we provide a taxonomy and detailed characterization of 25 architecturally significant MLOps guidelines for ML model integration and deployment. These guidelines are intended to support practitioners in taking the right approach when designing MLOps systems, specifically for integrating ML models and deploying ML models in their MLOps systems. The guidelines are derived from a gray literature review (GLR) of practitioner-driven evidence provided as MLOps practices online.
Among the results, we observe a strong emphasis on designing systems that make extensive use of automation in MLOps, as well as establishing mature CI/CD pipelines from the beginning of the MLOps projects. The sources also highlight practices such as standardized model export formats, which specifically facilitate the integration of new ML components into existing systems. The use of containerization is also the most advised approach to improve portability and reproducibility.
For future work, we plan to extend the description of the guidelines with their associated application context, trade-offs, and challenges, as an extension to our GLR. 
We also plan to further validate our guidelines with experienced practitioners in focus groups.
\\
\\
\textbf{Data Availability:} The data and artifacts of this study are available at~\cite{replication_package}.

\begin{credits}
\subsubsection{\ackname} 
This research is supported by ExtremeXP, a project co-funded by the European Union Horizon Programme under Grant Agreement No. 101093164.
\end{credits}

\renewcommand{\bibsection}{\section*{References}} 
\bibliographystyle{splncs04nat}
\begingroup
\small 
\bibliography{bibliography}
\endgroup

\end{document}